\documentclass[reprint,amsmath,amssymb,nofootinbib,aps,prd]{revtex4-1}

\usepackage{graphicx}% Include figure files
\usepackage{dcolumn}% Align table columns on decimal point
\usepackage{bm}% bold math
\usepackage{siunitx}

\usepackage{hhline}
\usepackage[american]{babel}
\usepackage{booktabs}
\usepackage{dcolumn}  % Align table columns on decimal point
\usepackage[T1]{fontenc}  % Font encoding that enables acents in the PDF
\usepackage[colorlinks]{hyperref}  % Add hypertext capabilities
\usepackage[utf8]{inputenc}  % Accept the Western European characters set
\usepackage{multirow}
\usepackage{txfonts}  % Fonts: Times (text) and Txfonts (math)
\usepackage{url}
\usepackage[dvipsnames]{xcolor}
\usepackage{float}

\usepackage{array}

\newcolumntype{d}[1]{D{.}{.}{#1}}
\newcolumntype{v}[1]{D{,}{,\ }{#1}}
\DeclareSIPrePower\quartic{4}

\usepackage{booktabs}
\AtBeginDocument{
	\heavyrulewidth=.08em
	\lightrulewidth=.05em
	\cmidrulewidth=.03em
	\belowrulesep=.65ex
	\belowbottomsep=0pt
	\aboverulesep=.4ex
	\abovetopsep=0pt
	\cmidrulesep=\doublerulesep
	\cmidrulekern=.5em
	\defaultaddspace=.5em
}

\begin{document}

%\preprint{APS/123-QED}

\title{Constraining nonlinear corrections to Maxwell electrodynamics using $\gamma\gamma$ scattering}

\author{P. Niau Akmansoy}
	\email{pniau7@gmail.com}
	\affiliation{Departamento de Física Teórica e Experimental,\\
				Universidade Federal do Rio Grande do Norte. Campus Universitário,
				s/n - Lagoa Nova, CEP 59072-970, Natal, Brazil}
\author{L. G. Medeiros}
	\email{leogmedeiros@ect.ufrn.br}
	\affiliation{Escola de Ciências e Tecnologia,\\
				Universidade Federal do Rio Grande do Norte. Campus Universitário,
				s/n - Lagoa Nova, CEP 59072-970, Natal, Brazil}

\date{\today}% It is always \today, today,
             %  but any date may be explicitly specified

\begin{abstract}
The recent light-by-light scattering cross section measurement made by the ATLAS\ Collaboration is used to constrain nonlinear corrections to Maxwell electrodynamics parametrized by the Lagrangian $L=F+4\alpha F^{2}+4\beta G^{2}+4\delta FG$. The ion's radiation is described using the equivalent photon approximation, and the influence of four different nuclear charge distributions is evaluated. Special attention is given to the interference term between the Standard Model and the nonlinear corrections amplitudes. By virtue of the quadratic dependence on $\alpha$, $\beta$ and $\delta$, the nonlinear contribution to the Standard Model $\gamma \gamma $ cross section is able to delimit a finite region of the parameter's phase space. The upper values for $\alpha$, $\beta$ in this region are of order $10^{-10}\si{\per\quartic\giga\electronvolt}$, a constraint of at least $12$ orders of magnitude more precise when compared to low-energy experiments. An upper value of the same order for $\delta$ is obtained for the first time in the LHC energy regime. We also give our predictions for the Standard Model cross section measured at ATLAS for each distribution and analyze the impact of the absorption factor. We finally give predictions for the future measurements to be done with upgraded tracking acceptance $\left\vert \eta \right\vert <4$ by the ATLAS Collaboration.
\end{abstract}

\maketitle

\section{\label{Introduction} Introduction}
Maxwell electrodynamics is one of the most successful theories in physics. Since its publication in 1873, it has been the source of notable predictions, such as electromagnetic waves, and served as a keystone for the proposal of new theories, such as Einstein's special relativity. The efforts to quantize the theory of electrodynamics helped to lay the foundations of quantum field theory. Its quantized version is capable of matching experimental results up to $10$ parts per billion \cite{Beringer} making it one of the most precise theories available. Despite all these achievements, the increasing ingenuity of new experiments, in both low- and high-energy domains, imposes the necessity to keep testing, whether to validate the theory or to find new sources of physics.

Historically, Maxwell's equations were derived phenomenologically. It is interesting, however, to look at them from another point of view. Following a "bottom-up" approach, its Lagrangian can be derived imposing a Lorentz invariant gauge theory with $U(1)$ symmetry and second-order linear equations of motion for the potentials \cite{Utiyama}. In this way, generalizations of Maxwell electrodynamics can be obtained by breaking at least one of the restrictions mentioned above. Indeed, Proca and Podolsky electrodynamics arise by breaking the internal $U(1)$ symmetry - introducing a mass term -, and allowing higher order equations of motion, respectively \cite{Proca,Bopp,Podolsky}. On the other hand, by allowing nonlinear equations of motion, an interesting class of electrodynamics, which are generically called nonlinear electrodynamics (NLED), arises \cite{Plebanski}. The most well-known examples of NLED are Euler-Heisenberg \cite{EulerHeis} and Born-Infeld \cite{BornInfeld1,BornInfeld2} theories, both proposed in the $1930$s with very different purposes. The first one emerges as a direct consequence of Dirac's relativistic theory of the electron, while the second arises as an attempt to solve the divergence of a pointlike particle potential. It is noteworthy that interest in Born-Infeld theory was revived after it was shown that it arises as the underlying electrodynamics in the low-energy regime of string theories \cite{Fradkin1985}.

In the present paper, we focus our study on nonlinear corrections to Maxwell electrodynamics, which includes NLED in regimes where their Lagrangians can be correctly described by the first terms of their respective MacLaurin series. Consequences of these corrections are well known and are expected if QED proves to be right \cite{Rizzo}. For this reason, several groups are currently working on proposing feasible tests based on these phenomena. Low-energy experiments, such as PVLAS \cite{PVLAS} and BMV \cite{BMV}, are built to detect the presence of magnetic birefringence by measuring the ellipticity acquired by a linearly polarized beam after traversing a magnetic field. While their current results are compatible with zero, they can be used to restrict a region of the parameter space constraining nonlinear corrections, such as was done in \cite{Fouche}. These experiments, however, are sensitive to specific combinations of the parameters, and thus cannot completely constrain the phase space by themselves.

The hydrogen atom - and more generally hydrogen-like atoms - form a neat low-energy laboratory to test for nonlinear corrections. High precision measurements of their transition energies are readily found in the literature \cite{Kramida}. Through perturbation theory, it is possible to analyze the modification of the energy spectrum by the inclusion of several terms in the Lagrangian. In particular, this framework can be used to study how the modification of Coulomb's potential due to NLED affects the ground-state energy. Comparison with experimental results constrains the magnitude of these corrections and, consequently, the parameters of the theory \cite{Soff,Carley,AtomHydr}. It is noteworthy that the complete Lagrangian is needed in this procedure, which imposes a particular analysis for each theory.

The equations of motion that describe classical electrodynamics are linear and, as such, cannot predict the interaction between electromagnetic waves in vacuum. Light-by-light scattering is a purely quantum process, which arises as a consequence of vacuum polarization and occurs at leading-order via an $\mathcal{O}\left(\alpha^4\right)$ virtual one-loop diagram consisting of charged particles. This phenomenon has already been measured indirectly through the electron's and muon's anomalous magnetic moment \cite{Dyck,Brown}. Recently, in $2013$, d'Enterria and Silveira suggested that the observation of light-by-light scattering would be achievable at LHC energies in ultraperipheral collisions with heavy-ions \cite{dEnterria}. As a consequence, the ATLAS Collaboration announced the first direct detection in $2016$ \cite{ATLAS}.

If the vacuum is invariant by $C$, $P$, and $T$ transformations, the first-order nonlinear corrections can be described by the addition of Lorentz invariants $F^2$ and $G^2$ to Maxwell's Lagrangian [see Eqs. \eqref{F2} and \eqref{G2} for definitions]. However, if we allow $CP$ violation, the term $FG$ must also be added. Contributions to such a term may come from within the Standard Model, from the weak and strong sectors \cite{Millo2009}, or from beyond Standard Model physics \cite{Liao2007}. When compared to the free Lagrangian, these terms dominate at high-energy regimes where their effects become relevant. For this reason, the light-by-light scattering cross section may be used to obtain today's most precise constraints for nonlinear corrections to Maxwell electrodynamics. This idea has already been used to constrain Born-Infeld's parameter \cite{Ellis}.

In this work, we completely constrain the phase space of nonlinear parameters associated with the $F^2$, $G^2$ and $FG$ terms. Using the equivalent photon approximation, we compare the results obtained with four different charge distributions and study the impact of the absorption factor \cite{Baur,Jackson}, and the relevance of the interference term arising between nonlinear corrections and the Standard Model amplitudes.

This paper is organized in the following way. In Sec. \ref{NLCorrections}, imposing a series of requirements, the general form for the Lagrangian describing nonlinear corrections to Maxwell electrodynamics is presented and followed by a brief discussion of its consequences. Considering corrections up to quadratic order in the invariants, we deduce the differential and total cross sections for the elastic nonpolarized $\gamma\gamma$ scattering. In Sec. \ref{EPA}, we detail the necessary ingredients for the theoretical description of the experiment. Four different distributions are proposed to describe the nuclear charge. Using the equivalent photon approximation, the ions are then treated as high-energy photon sources. We also review the experimental cuts and describe the $\gamma\gamma$ cross section according to the Standard Model and nonlinear corrections. Next, in Sec. \ref{Results}, we present our calculations based on the Standard Model for the cross section measured at ATLAS. We compare the results obtained from each charge distribution, with and without the absorption factor. Using the cross section measured by the ATLAS Collaboration, we derive an expression which fully constrains the phase space accessible to the parameters. Finally, we end the section with our prediction for the cross section to be measured at LHC with extended acceptance tracking. Our conclusions are given in Sec. \ref{Conclusion}.

\section{\label{NLCorrections} Nonlinear Corrections}
There are several approaches available in physics which allows one to calculate observables to any accuracy desired. When the full theory is known, we are able to make predictions at any energy scale. However, when some degrees of freedom of the theory are large compared to the scale of interest, it is often appropriate to integrate them out. This top-down approach is used in order to obtain a simpler description of the relevant phenomena in a particular energy regime. An example of such is Euler-Heisenberg theory when the energies are much smaller than the mass of the electron, $m_e$.

On the other hand, it is possible that the full theory is not known, is nonperturbative in the scale of interest, or even exist. In this case, an effective theory can be built by writing down a Lagrangian with all possible operators following a set of rules and symmetries that the theory should satisfy in the energy regime of interest. This bottom-up approach has been used in several areas of physics. The most well-known example of this approach is the beta decay theory proposed by Fermi when only the hadrons and leptons undergoing weak decay were known. Although Fermi's interaction is nonrenomalizable and violates unitarity at high-energies, it was able to correctly describe the process in the low-energy regime.

Following a bottom-up effective field theory approach, the form of the Lagrangian for a generic nonlinear electrodynamics theory is greatly restricted by the imposition of the Lorentz and gauge group symmetries. The only relativistic gauge invariants available are $F$ and $G$ which can be defined as
\begin{align}
		F &\equiv -\frac{1}{4}F^{\mu \nu }F_{\mu \nu }=\frac{1}{2}\left(
		E^{2}-B^{2}\right),\label{F2} \\
		G &\equiv -\frac{1}{4}\tilde{F}^{\mu \nu }F_{\mu \nu }=\vec{E}\cdot \vec{B},\label{G2}
\end{align}
where $F^{\mu \nu }$ is the electromagnetic field strength and $\tilde{F}^{\mu \nu }\equiv\frac{1}{2}\varepsilon^{\mu \nu \alpha \beta }F_{\alpha\beta }$ its dual.  We focus on nonlinear corrections coming from theories of which the Lagrangian are expressible through analytic functions. In this way, theories can be described as a MacLaurin series in the invariants \cite{Rizzo},
\begin{equation}
	L=\sum_{i,j=0}^{\infty }c_{ij}F^{i}G^{j}.
\end{equation}
Since the features of these theories arise in intense field regimes, all of them must be, in the weak field limit, indistinguishable from Maxwell electrodynamics. Thus, the coefficients must be chosen to be $c_{00}=c_{01}=0$ and $c_{10}=1$. The first terms of the Lagrangian expansion are then
\begin{equation} \label{LagSerE}
L=F+c_{20}F^{2}+c_{02}G^{2}+c_{11}FG+...,
\end{equation}%
where the first term is Maxwell's Lagrangian. Due to the analyticity of the Lagrangians, their power series must always converge inside a convergence radius or energy regime. For this to be true, below some characteristic energy scale $\Lambda$, each term must consistently be less relevant when compared with the ones with a lower degree. Thus, any NLED that satisfies all previous requirements can be described by \eqref{LagSerE}. Theories in which $\Lambda$ is much greater than the energies involved in the LHC, their Lagrangian can be correctly approximated by the first terms of the series.

For the purposes of this investigation, we consider as relevant terms up to second order in the invariants,
\begin{equation} \label{LagSerT}
L=F+4\alpha F^{2}+4\beta G^{2}+4\delta FG,
\end{equation}%
where $\alpha $, $\beta $ and $\delta$ are parameters with dimension of energy to the inverse fourth power. It is important to remember that each nonlinear theory possesses a particular energy regime at which their effects become relevant and thus, the validity of the expansion \eqref{LagSerT} needs to be verified for each one them separately. It is possible to study from \eqref{LagSerT} the behavior of several theories by simply matching the coefficients. As an example, to recover Born-Infeld theory - and in general, Born-Infeld-like theories \cite{Helayel1,Helayel2,Kruglov}- we must choose $\alpha =\beta =\frac{1}{8b^{2}}$ and $\delta=0$, where $b$ represents the maximum value of the electric field. \footnote{This maximum value for the electric field may vary from one Born-Infeld-like theory to another and may not even exist, such as in the case of the exponential electrodynamic.}

The presence of these nonlinear corrections has profound consequences. Classically, they may be interpreted, through the classical constitutive equations, as giving rise to dielectric properties of the vacuum. The electric permittivity and the magnetic permeability are now tensors and depend on the electromagnetic field itself. As a result of this, several nonlinear processes emerge in the presence of an external electromagnetic field. From this point of view, Euler-Heisenberg effective theory classically describes the effects of vacuum polarization due to electron-positron pair creation in energy regimes well below the electron's mass \cite{Rizzo}.

On the other hand, the quantization of \eqref{LagSerE} gives rise to the direct autointeraction of photons without resorting to any intermediate virtual particle. The interaction between four photons can be made explicit by rewriting the Lagrangian \eqref{LagSerT} as
\begin{multline} \label{LagGamma}
	L=L_{0} +\gamma^{\left[ A_{12}A_{34}\right] \left[ A_{56}A_{78}\right]}
	\left(\partial _{a_{1}}A_{a_{2}} \right)
	\left(\partial_{a_{3}}A_{a_{4}} \right)
    \left(\partial _{a_{5}}A_{a_{6}} \right)
	\left(\partial _{a_{7}}A_{a_{8}} \right) ,
\end{multline}%
where $L_{0}$ is Maxwell's Lagrangian, $\partial _{a_{1}}A_{a_{2}}$ is the derivative of the $4$-potential contracted with the eight-dimensional matrix $\gamma^{\left[A_{12}A_{34}\right]\left[ A_{56}A_{78}\right]}$. The gamma matrix is defined as
\begin{align}
\begin{split}
\label{Gamma}
\gamma^{\left[ A_{12}A_{34}\right] \left[ A_{56}A_{78}\right]}
\equiv& \alpha \gamma_{F^{2}}^{\left[A_{12}A_{34}\right] \left[A_{56}A_{78}\right]}\\
&+\beta \gamma_{G^{2}}^{\left[A_{12}A_{34}\right] \left[A_{56}A_{78}\right]}
+\delta \gamma_{FG}^{\left[A_{12}A_{34}\right] \left[A_{56}A_{78}\right]},
\end{split}
\end{align}
where $\alpha $ and $\beta $ are the same parameters found in \eqref{LagSerT}, and with
\begin{align}
\begin{split}
	\gamma_{F^{2}}^{\left[ A_{12}A_{34}\right]\left[ A_{56}A_{78}\right] }
	&\equiv \delta ^{a_{1}a_{3}}\delta ^{a_{2}a_{4}}\delta ^{a_{5}a_{7}}\delta
	^{a_{6}a_{8}}
	\\
	-2\delta ^{a_{1}a_{3}}&\delta ^{a_{2}a_{4}}\delta
	^{a_{5}a_{8}}\delta ^{a_{6}a_{7}}+\delta ^{a_{1}a_{4}}\delta
	^{a_{2}a_{3}}\delta ^{a_{5}a_{8}}\delta ^{a_{6}a_{7}}, 
\end{split}
\end{align}
\begin{equation}
	\gamma _{G^{2}}^{\left[ A_{12}A_{34}\right] \left[ A_{56}A_{78}\right] }
	\equiv \varepsilon ^{a_{1}a_{2}a_{3}a_{4}}\varepsilon
	^{a_{5}a_{6}a_{7}a_{8}},
\end{equation}
and
\begin{equation}
\gamma _{FG}^{\left[ A_{12}A_{34}\right] \left[ A_{56}A_{78}\right] }
\equiv -\left(\delta ^{a_{1}a_{3}}\delta ^{a_{2}a_{4}}-\delta ^{a_{1}a_{4}}\delta ^{a_{2}a_{3}}\right)\varepsilon^{a_{5}a_{6}a_{7}a_{8}}.
\end{equation}
We have used a block notation $A_{ij}\equiv a_{i}a_{j}$ to emphasize the matrices' invariance through their permutation. For example, permuting $A_{12}\leftrightarrow A_{34}$ indicates that we need to permute $a_{1}\leftrightarrow $ $a_{3}$ and $a_{2}\leftrightarrow $ $a_{4}$ simultaneously. The matrix $\gamma _{G^{2}}$, is also symmetric by the simultaneous permutation of $A_{12}\leftrightarrow A_{56}$ and $A_{34}\leftrightarrow A_{78}$. With the help of these properties, we are able to derive the probability amplitude for the elastic $\gamma \gamma \rightarrow \gamma \gamma$ scattering, which can be written as
\begin{align} \label{Ampl}
\begin{split}
	\mathcal{M}_{NL}  =&
	\left[ P_{A_{12}A_{34}A_{56}A_{78}}\gamma ^{\left[
		A_{12}A_{34}\right] \left[ A_{56}A_{78}\right] }\right]
	\\
	 &\times p_{a_{1}}\varepsilon _{a_{2}}^{\ast }\left( p\right)
	p_{a_{3}}^{\prime }\varepsilon _{a_{4}}^{\ast }\left( p^{\prime }\right)
	k_{a_{5}}\varepsilon _{a_{6}}\left( m\right) k_{a_{7}}^{\prime
	}\varepsilon _{a_{8}}\left( k^{\prime }\right) ,
\end{split}
\end{align}
where $p$ and $\varepsilon \left( p,i\right) $ generically represents the $4$-momentum and the polarization vector of the photons. In \eqref{Ampl}, the totally symmetric permutation operator $P_{A_{12}A_{34}A_{56}A_{78}}$ acts on $\gamma$ and indicates that we must add together all possible $\gamma$s with permutated indices. As should be expected, the substitution of \eqref{Gamma} into \eqref{Ampl} gives the total amplitude as the sum of the amplitudes due to each of the squared invariant terms and thus is a linear function of the parameters $\alpha $, $\beta $ and $\delta$. As a result, the parameters can be easily extracted from the interference term between \eqref{Ampl} and the leading-order amplitude from the Standard Model, easing up numerical computations.

The nonpolarized square of \eqref{Ampl} can be expressed in a simple and reference-independent way in terms of Mandelstam's variables as:
\begin{equation}
\frac{1}{4}\sum_{\text{Pol.}}\left\vert \mathcal{M}\right\vert ^{2}=4\left[ \frac{1}{2}%
\left( \alpha -\beta \right) ^{2}+\left( \alpha ^{2}+\beta ^{2}+\delta^2\right) %
\right] \left( s^{4}+t^{4}+u^{4}\right) .  \label{AmplQuad}
\end{equation}
Particularizing to the center-of-mass frame allows us to write the differential cross section as
\begin{equation}
\left( \frac{d\sigma }{d\Omega }\right) _{CM}=\left[ 
\frac{1}{2}\left( \alpha -\beta \right) ^{2}+\left( \alpha ^{2}+\beta^{2}+\delta^2\right) \right]
\frac{\left( \cos 2\theta +7\right) ^{2}m_{\gamma \gamma }^{6}}{512\pi ^{2}} ,
\label{SecChoqDif}
\end{equation}
and the total cross section as
\begin{equation}
\sigma _{CM}=\frac{7}{40\pi }\left[ \frac{1}{2}\left( \alpha -\beta \right)
^{2}+\left( \alpha ^{2}+\beta ^{2}+\delta^2\right) \right] m_{\gamma \gamma }^{6},
\label{SecChoqTot}
\end{equation}
where $m_{\gamma \gamma }$ is the total energy in center-of-momentum frame or the invariant mass of the diphoton system. A similar result has been obtained in Refs. \cite{Fichet,Rebhan2017}. It is interesting to notice the lack of symmetry of the parameter $\delta$, when compared to $\alpha$ and $\beta$, in Eq. \eqref{SecChoqTot}. This is due to the fact that the unpolarized interference term between the $CP$-odd and $CP$-even terms is zero. From a dimensional point of view, the cross section's dependence on the sixth power of the invariant mass can be expected from the linear dependence of \eqref{Ampl} on the parameters. Furthermore, this dependence, which is characteristic of an effective field theory, will violate both unitarity and the so-called Froissart bound \cite{Froissart1961} - which limits the growth of the total cross section to approximately $\log^2 m_{\gamma\gamma}$ - outside of the valid energy regime. By matching the coefficients, it is possible to recover known results for Born-Infeld and Heisenberg-Euler, obtaining $\alpha _{BI}=\beta _{BI}=\frac{1}{8b^{2}}$ and $\delta_{BI}=0$, and $\alpha _{HE}=\frac{4}{7}\beta _{HE}=\frac{4}{90}\frac{\alpha ^{2}}{m^{4}}$ and $\delta_{HE}=0$, respectively \cite{Itzykson,Schubert}.

\section{\label{EPA} $\gamma\gamma$ scattering in the equivalent photon approximation}

\begin{figure}[h!]
	%\vspace{-5mm}
	\centering
	\includegraphics[scale=0.7]{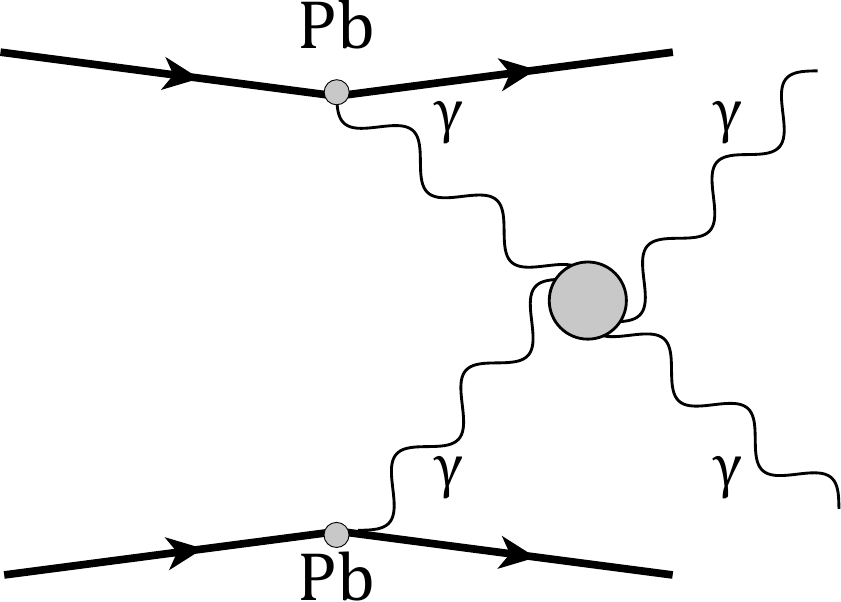}
	%\vspace{-0.6 cm}
	\caption[Ultraperipheral collision of lead ions]{\footnotesize Ultraperipheral collision of lead ions. Charged particles scattering with impact-parameter greater than the sum of their radii. Quasivirtual photons emitted by the ions scatter producing a new pair of photons.}
	\label{UPC}
\end{figure}

The LHC has been optimized for proton collisions and therefore most of the physics coming from it is based on that kind of experiment. However, for a short period of the year - for one or two months - the LHC is dedicated to heavy-ion collisions. Analyzing $480\mu b^{-1}$ of lead-$208$-ion collision data collected in $2015$, the ATLAS Collaboration announced the detection of light-by-light scattering with a cross section of $70\pm 24$(stat.)$\pm 17$(syst.) $\si{\nano\barn}$ \footnote{Most of the systematic uncertainty comes from photon reconstruction and identification efficiency uncertainties.} \cite{ATLAS}. This process can be produced in ultraperipheral collisions (UPCs) of charged particles, where they cross each other with an impact parameter greater than the sum of the ion's radii (Fig. \ref{UPC}). This kind of collision has the advantage of avoiding strong interaction from nuclear overlap and thus cleaning the signal's background.

%\newpage

Any charged particle accelerated at high-energies produces an intense electromagnetic field \cite{JacksonBook}. Comparison of the electromagnetic energy flux with the photon flux in the frequency space allows estimating the distribution of photons emitted by the ion. This is the essence behind the semiclassical equivalent photon approximation (EPA) \cite{Weiz,Williams,Fermi}. In this way, charged particles taking part in electromagnetic processes can be replaced by their respective photon distribution. As observed by d'Enterria and Silveira \cite{dEnterria}, heavy ions allow observing light-by-light scattering due to the coherent production of radiation by their nucleons. The luminosity is enhanced by a factor of $Z^{4}\sim 10^{7}$ compensating the low cross section of order $O\left(\alpha ^{4}\right) \sim 10^{-9}$. On the other hand, the electromagnetic radiation produced by the nuclear charge distribution interferes destructively when wavelengths are of the order of the ion's radius $R$. This limits the upper value of the energy spectrum to $\omega_{\max }\approx\frac{\gamma}{R}\approx 80 \si{\giga\electronvolt}$ for the lead-$208$ ion \cite{dEnterria}.

In the EPA, the production of photons by the quasielastic scattering of ions in UPCs $AA\rightarrow A^{\ast }A^{\ast }+\gamma \gamma $ can be described by convoluting the subsystem $\gamma\gamma \rightarrow \gamma\gamma$ cross section with the effective photon flux \cite{Klusek2016}:
\begin{equation}  \label{Convo}
\sigma _{PbPb\rightarrow PbPb\gamma \gamma }=\int \sigma _{\gamma \gamma
	\rightarrow \gamma \gamma }\left( m_{\gamma \gamma }\right) dn_{\gamma
	\gamma },
\end{equation}
with
\begin{equation}
	\frac{dn_{\gamma \gamma }}{dbd^{2}b_{c}dm_{\gamma \gamma }dY}=\pi m_{\gamma \gamma }bN\left( \frac{m_{\gamma \gamma }}{%
		2}e^{Y},b_{1}\right) N\left( \frac{m_{\gamma \gamma }}{2}e^{-Y},b_{2}\right)S^{2}\left( b\right),
\end{equation}
where $m_{\gamma \gamma }=\sqrt{4\omega _{1}\omega _{2}}$ is the diphoton invariant mass, $Y=\frac{1}{2}\left(\eta_{1}+\eta_{2}\right)$ is the rapidity of the diphoton in the lab reference, and $b_{1}$ and $b_{2}$ are the impact-parameters \footnote{The labels $1$ and $2$ refer to the ions and the photons each one produces.}. $N\left( \omega,b\right) $ represents the flux of photons with energy $\omega $ emitted by the ion at a distance $b$ in the plane perpendicular to the motion. The absorption factor $S^{2}\left( b\right)$ encodes the ion's probability of survival when scattering with impact parameter $b$, ensuring that only UPCs are considered and can be conveniently described in a first approximation as 
\begin{equation}
S^{2}\left( b\right) =\Theta \left( b-2R\right) ,  \label{FatAbs}
\end{equation}%
where $R\simeq 7.1 \si{\femto\meter}$ is the lead radius. The connection between the impact-parameters is given by the expressions:
\begin{equation*}
	\vec{b}=\vec{b}_{1}-\vec{b}_{2}\text{ \ \ and \ \ }\vec{b}_{c}=\frac{\vec{b}%
		_{1}+\vec{b}_{2}}{2}.
\end{equation*}%
More details on this framework can be found in Ref. \cite{Klusek2011}.

The photon flux in the impact parameter space can be written as:
\begin{equation}
	N\left( \omega ,b\right) =\frac{Z^{2}\alpha }{\pi }\frac{1}{\omega }\phi
	\left( \omega ,b\right) ^{2},
\end{equation}%
where the function $\phi \left( \omega ,b\right) $, given by 
\begin{equation}
	\phi \left( \omega ,b\right) =\int_{\frac{\omega }{\gamma }}^{\infty }\frac{1%
	}{u}\sqrt{u^{2}-\left( \frac{\omega }{\gamma }\right) ^{2}}J_{1}\left( b%
	\sqrt{u^{2}-\left( \frac{\omega }{\gamma }\right) ^{2}}\right) F\left(
	u\right) du,
\end{equation}
is associated with the intensity of the electric field produced by the ion, $\gamma$ is the ion's Lorentz factor, and $J_{1}\left( x\right) $ is the Bessel function of the first kind. The main ingredient of the photon flux is the form factor $F\left( u\right)$ given by the Fourier transform of the charge distribution:
\begin{equation}
F\left(q\right)=4\pi\int_{0}^{\infty}drr^2\rho\left(r\right)\frac{\sin\left(qr\right)}{qr},
\end{equation}
for radially symmetric charge distributions. There are several parameterization for the charge distribution, and they introduce an important theoretical uncertainty \cite{dEnterria}. More realistic, and therefore complex, parametrization carries more details of the charge distribution introducing proximity effects absent in others. However, at large impact parameters, all parametrizations should be equivalent.

\begin{table}[]
	\renewcommand {\arraystretch}{2}
	\setlength{\tabcolsep}{8pt}
	\centering
	\begin{tabular}{*3c}
		\toprule
		Model    & Charge distribution                                                & Form factor                               \\
		\midrule
		Yukawa   & $\frac{\Lambda ^{2}}{4\pi }\frac{e^{-\Lambda r}}{r}$               & $\frac{\Lambda ^{2}}{\Lambda ^{2}+q^{2}}$ \\
		Fermi 2P & $\frac{\rho _{0}}{1+e^{\frac{r-c}{a}}}$                            & \cite{Maximon}                                  \\
		Gaussian & $\frac{Q_{0}^{3}}{\sqrt{8\pi ^{2}}}e^{-\frac{1}{2}Q_{0}^{2}r^{2}}$ & $e^{-\frac{1}{2}\frac{q^{2}}{Q_{0}^{2}}}$ \\
		Sphere   & $\frac{3}{4\pi R^{3}}\Theta \left( R-r\right) $                    & $\frac{3j_{1}\left( qR\right)}{qR}$\\
		\bottomrule
	\end{tabular}
	\caption[Charge distributions and corresponding form factos]{\footnotesize Charge distributions and corresponding form factors. Yukawa and Gaussian distribution parameters, $\Lambda =0.088\si{\giga\electronvolt}$ and $Q_{0}=0.060\si{\giga\electronvolt}$, are such as to obtain the lead root-mean-square radius \cite{Hencken,Baur}. Fermi 2P distribution parameters $a=0.549\si{\femto\meter}$ and $c=6.642\si{\femto\meter}$ describe the diffuseness and the radius of the lead ion respectively \cite{Burleson}. The homogeneously charged sphere distribution is characterized by the ion's radius $R=7.1\si{\femto\meter}$ \cite{Baur2}.}
	\label{TabFatForm}
\end{table}

\begin{figure}[h!]
	%\vspace{-5mm}
	\centering
	\includegraphics[scale=0.9]{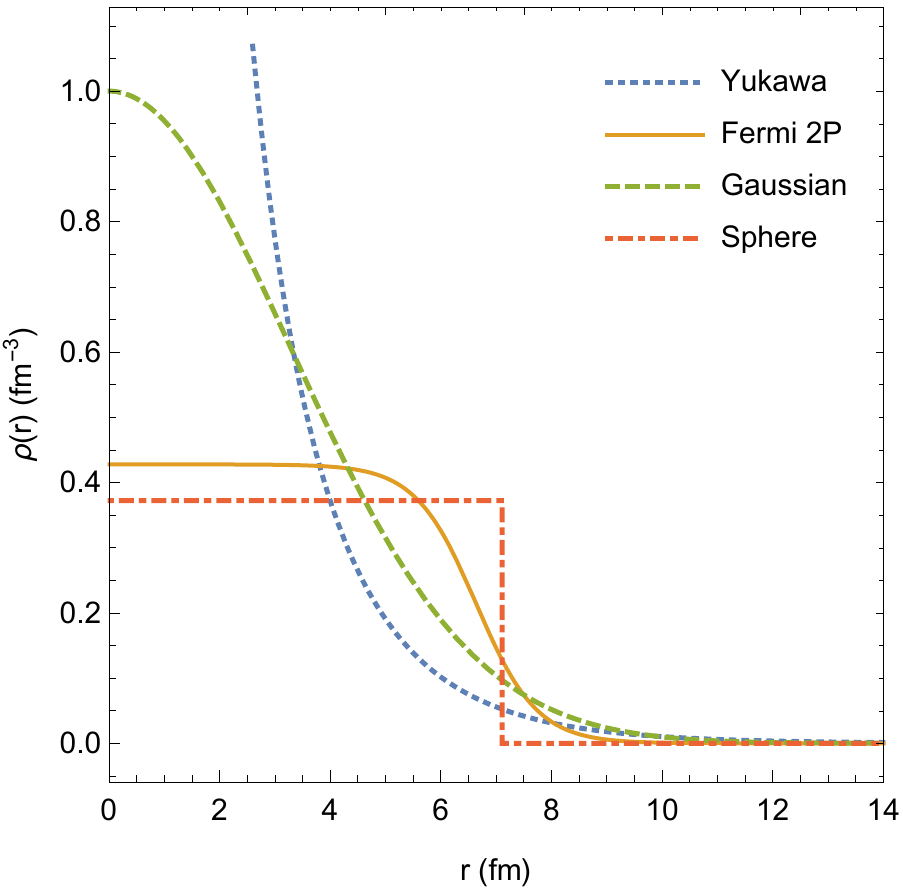}
	%\vspace{-0.6 cm}
	\caption[Normalized nuclear charge distributions]{\footnotesize Plot of normalized nuclear charge distributions given in Table \ref{TabFatForm}}
		\label{DistrCargPlot}
	\end{figure}

We compare the results obtained with the four charge distributions shown in Table \ref{TabFatForm}. While a Yukawa charge distribution is considered rather unrealistic, it has the advantage of allowing an analytical expression for $\phi \left( \omega ,b\right) $,
\begin{equation}
	\phi \left( \omega ,b\right) =\frac{\omega }{\gamma }K_{1}\left( \frac{%
		b\omega }{\gamma }\right) -\sqrt{\left( \frac{\omega }{\gamma }\right)
		^{2}+\Lambda ^{2}}K_{1}\left( b\sqrt{\left( \frac{\omega }{\gamma }\right)
		^{2}+\Lambda ^{2}}\right) .
\end{equation}
A second charge distribution is parametrized using a Fermi with two parameters (2P) model \cite{Burleson}. The constant $\rho _{0}$ is such that its form factor is normalized to $1$ at the origin. This model is considered much more realistic but has no closed form for its corresponding form factor. An expression, however, can be obtained in terms of a series \cite{Maximon}. Two other widely used distributions in the literature, Gaussian and of a homogeneously charged sphere, have simple form factors and are included for comparison \cite{Hencken}. A normalized plot of these charge distributions is shown in Fig. \ref{DistrCargPlot}.

%\newpage

Several triggers and cuts are used during the operation of the detector and along the analysis of the data. At the LHC, up to $40$ million collisions per second can occur, each one producing several events. Triggers are chosen in order to reduce this huge amount of information to the roughly $400$ events per seconds that the ATLAS detector is capable of recording. Then, cuts are applied to clean the signal from its background and to take advantage of the detector's components efficiencies. As a consequence, in order to correctly predict the measurements, this cuts must be included. During the analysis of the $\gamma \gamma$ scattering, the main cuts used by the ATLAS Collaboration to select events were individual photon transverse momentum $p_{t}>3 \si{\giga\electronvolt}$, pseudorapidity $\left\vert \eta \right\vert <2.37$ (excluding the electromagnetic calorimeter transition region $1.37<\left\vert \eta \right\vert <1.52$) and invariant diphoton mass $m_{\gamma \gamma }>6\si{\giga\electronvolt}$. To include these cuts, we replace the total cross section with the differential distribution $\sigma _{\gamma \gamma }\rightarrow \int \frac{d\sigma _{\gamma \gamma }}{dp_{t}}dp_{t}$, and perform a change of variables transforming the invariant mass $m_{\gamma \gamma }$ and the diphoton rapidity $Y$ into the rapidities of the outgoing photons $\eta_{1}$ and $\eta_{2}$ using
\begin{align}  \label{VarChange}
m_{\gamma \gamma }&=2p_{t}\cosh \left( \frac{\eta_{1}-\eta_{2}}{2}\right), \\ 
Y&=\frac{1}{2}\left(\eta_{1}+\eta_{2}\right).
\end{align}
Photons produced by the ions may have transverse momentum up to $q_{\bot}\simeq 1/R\simeq 28\si{\mega\electronvolt}$ but are assumed to be emitted along the beam in order to derive \eqref{VarChange}. This assumption is part of the EPA scheme and connects the center-of-mass reference frame to the laboratory frame through a simple boost in the $z$-direction. Other cuts, such as on the diphoton transverse momentum $p_{t}^{\gamma \gamma }<2\si{\giga\electronvolt}$ and on the acoplanarity $1-\Delta \phi _{\gamma \gamma }/\pi <0.01$, are imposed as a total fixed cut of $15\%$ estimated using Table $1$ in Ref. \cite{ATLAS}.

\begin{figure}[h!]
	%\vspace{-5mm}
	\centering
	\includegraphics[scale=0.65]{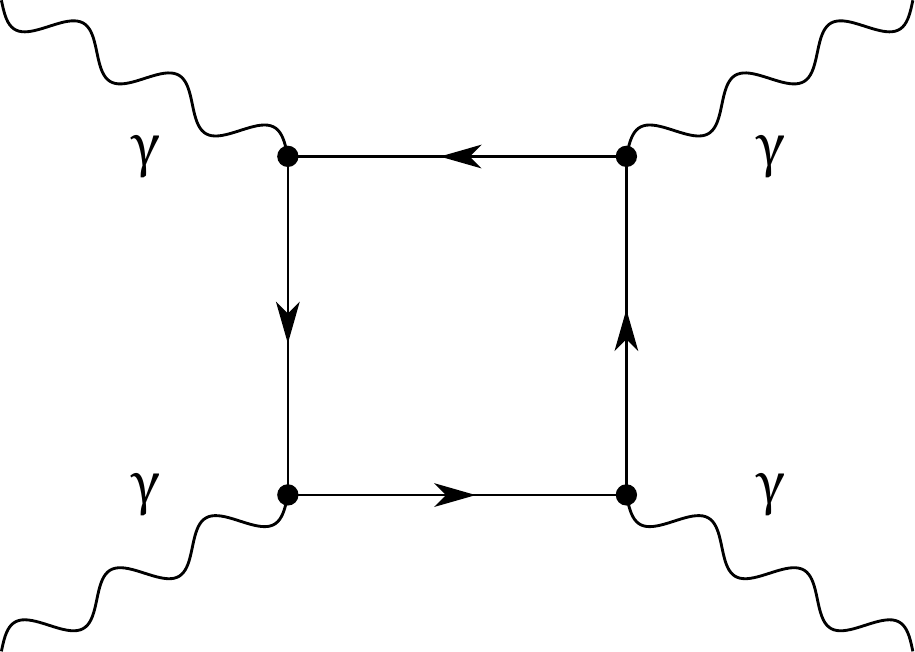}
	%\vspace{-0.6 cm}
	\caption[SM Feynman diagram]{\footnotesize Leading-order diagram for the Standard Model $\gamma\gamma $ scattering. Leptons, quarks and bosons $W^{\pm }$ are the main particles composing the loop.}
	\label{diagramaFeyn}
\end{figure}%\newpage

The $\gamma \gamma \rightarrow \gamma \gamma $ scattering has contributions from several mechanisms. In the Standard Model, the leading-order contribution proceeds via virtual one-loop box diagram (Fig. \ref{diagramaFeyn}). The elementary particles that can compose the loop are charged fermions (leptons and quarks) and bosons ($W^{\pm}$). The main contribution from each one of those particles is at energies around three times their masses. Thus, for the LHC energy regime and the physical limitations established by the ion's charge distribution, contributions coming from the $W^{\pm}$ bosons and $t$ quark are negligible. Another process by which photons fluctuate into vector mesons, called the vector-meson dominance-Regge mechanism, has contributions in the experiment energy range. However, the photons produced by this mechanism are very forwarded ($\left\vert \eta \right\vert \simeq 5$ or $\theta \simeq \ang{0.77}$ with the beam) and cannot be detected by the ATLAS detector. Furthermore, their transverse momentum is such that applied cuts would completely kill their contribution \cite{Klusek2016}. Therefore, only leptons and light quarks are considered. It is also worth mentioning that the QED and QCD next-to-leading-order corrections amount to approximately $0.35\%$ and $3\%$, respectively, when compared to the leading order of the photon-photon cross section in the ultrarelativistic limit \cite{Bern2001}.

\begin{figure}[h!]
	%\vspace{-5mm}
	\centering
	\includegraphics[scale=0.85]{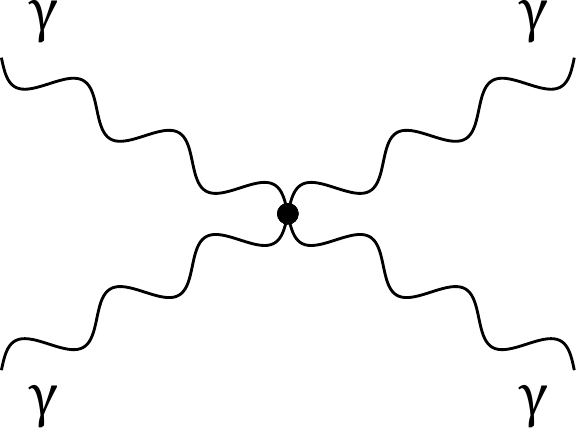}
	%\vspace{-0.6 cm}
	\caption[NL Feynman diagram]{\footnotesize Interaction vertex due to nonlinear correction terms $F^{2}$, $G^{2}$ and $FG$.}
	\label{diagramaFeynNL}
\end{figure}

Extensions of the Standard Model introduce all sorts of contributions through hypothetical charged or neutral virtual particles. In this sense, the light-by-light scattering can be used as a way to probe the quantum vacuum. Nonlinear electrodynamics introduce interaction vertices allowing photon fusion. In particular, the lowest-order correction terms $F^{2}$, $G^{2}$ and $FG$ allow direct interaction between four photons, as depicted in Fig. \ref{diagramaFeynNL}. The amplitude and cross sections due to these terms are given as a function of the parameters $\alpha $, $\beta $ and $\delta$ by Eqs. \eqref{Ampl}, \eqref{SecChoqDif} and \eqref{SecChoqTot}. It is noteworthy that, in the nonpolarized case, it is not possible to distinguish the contribution from $F^{2}$ and $G^{2}$.

%\newpage

In order to constrain the contribution to the total $\gamma \gamma $ cross section from nonlinear corrections, we assume the total amplitude of the process to be the sum of the Standard Model mechanisms mentioned above and (\ref{Ampl}). Hence, the theoretical cross section, to be compared with ATLAS' result, is composed of pure contributions due to the Standard Model and nonlinear corrections, and an interference term \footnote{In this case, the symbol $\cong $ means that both sides must be compatible.}:
\begin{equation} \label{TeoVsExp1}
\sigma _{SM}+\sigma _{NL}+\sigma _{I}\cong \sigma _{ATLAS}.
\end{equation}
As discussed in the previous section, the interference term is a linear function of $\alpha $, $\beta $ and $\delta$. Therefore, writing them out explicitly, Eq. \eqref{TeoVsExp1} can be rewritten as
\begin{align}
\begin{split}
\label{TeoVsExp2}	
\left[\frac{1}{2}\left( \alpha -\beta \right)^{2}+\left(\alpha ^{2}+\beta^{2}+\delta^2\right) \right] \bar{\sigma}_{NL}
+\underset{\sigma _{I}}{\underbrace{\alpha \bar{\sigma}_{\alpha }+\beta \bar{\sigma}_{\beta }}}&
\\
\cong \sigma_{ATLAS}&-\sigma _{SM},
\end{split}
\end{align}
where $\bar{\sigma}_{NL}$, $\bar{\sigma}_{\alpha }$ and $\bar{\sigma}_{\beta}$ are given in Table \ref{Tab2}, $\sigma _{ATLAS}=70\pm 24$(stat.)$\pm 17$(syst.) $\si{\nano\barn}$ and $\sigma _{SM}$ is given in Table \ref{Tab} for each distribution. It is noteworthy that the expected interference term $\bar{\sigma}_{\delta}=0$, just as was the case between \textit{CP}-odd and \textit{CP}-even interference term in Eq. \eqref{SecChoqTot}.

The Standard Model cross section was obtained using \texttt{FeynArts}3.10 \cite{FeynArts} to generate the diagrams and build the amplitude, \texttt{FormCalc}9.6 \cite{FCalcLTools} for algebraic simplifications and numerical computations, and \texttt{LoopTools}2.15 \cite{FCalcLTools} for loop calculations. For the interference term, we also used \texttt{FeynRules}2.3 \cite{FeynRules} package. Numerical results for purely nonlinear corrections were confronted with those obtained in Sec. \ref{NLCorrections}.

\section{\label{Results} Results}
In Table \ref{Tab} we list the result of our calculations using the Standard Model for the cross section measured at ATLAS. The results were obtained for each one of the four distributions presented in Sec. \ref{EPA}. Besides, for comparison purposes, we include the corresponding cross sections obtained by neglecting the absorption factor \eqref{FatAbs}. As a consequence, without the absorption factor, the integration over a wider range of the phase space overestimates the cross section by around $20\%$. \footnote{Strong interaction due to nuclear overlap was not taken into account.} It is an interesting fact that the cross sections obtained with the Gaussian and homogeneously charged sphere distributions differ from the one derived using Fermi 2P by less than $0.1\%$. On the other hand, cross sections obtained with Yukawa distribution are, in every case, $10\%$ larger than those obtained with Fermi 2P, in agreement with Ref. \cite{Klusek2016}. Theoretical uncertainties are mainly due to lack of knowledge in the ion's charge and are considered to be of order $20\%$ of the total cross section \cite{dEnterria}.

\begin{table}[]
	\setlength{\tabcolsep}{4pt}
	\renewcommand {\arraystretch}{1.5}
	\centering
	\begin{tabular}{*4c}
	\toprule
		Model                      & $\bar{\sigma}_{NL}$ ($\si{\giga\electronvolt}^{6}$) & $\bar{\sigma}_{\alpha }$ ($\si{\giga\electronvolt}^{2}$) & $\bar{\sigma}_{\beta }$ ($\si{\giga\electronvolt}^{2}$) \\
	\midrule
		Yukawa                     & $3.2\times 10^{21}$                                 & $-4.9\times 10^{8}$                                      & $-1.1\times 10^{9}$                                     \\
		Fermi, Gaussian, sphere & $2.5\times 10^{21}$                                 & $-4.1\times 10^{8}$                                      & $-9.3\times 10^{8}$                                   	\\ 
		\bottomrule
	\end{tabular}
	\caption[Numerical Parameters]{\footnotesize Proportionality constants. Numerical proportionality constants for the nonlinear and interference cross sections. See Eqs. \eqref{TeoVsExp2} and \eqref{IneqComp}.}
	\label{Tab2}
\end{table}

To constrain the parameters $\alpha $, $\beta $ and $\delta$, we deduct the Standard Model cross section prediction $\sigma_{SM}$ (second column of Table \ref{Tab}) from the experimental result obtained by the ATLAS Collaboration $\sigma_{ATLAS}$ and treat the remaining value as being produced by the nonlinear corrections alone (see Eq. \eqref{TeoVsExp2}). The theoretical, statistical and systematic uncertainties are added in quadrature. Using $3\sigma $ of confidence level, we are able to impose an upper limit on the nonlinear correction contribution given by the expression:
\begin{align}
 \label{IneqComp}
 \begin{split}
\frac{3}{2}\bar{\sigma}_{NL}\alpha ^{2}-\bar{\sigma}_{NL}\alpha \beta +\frac{3}{2}\bar{\sigma}_{NL}\beta ^{2}+&\bar{\sigma}_{NL}\delta^2+\bar{\sigma}_{\alpha }\alpha+\bar{\sigma}_{\beta }\beta
\\
&\leq\left\{\begin{tabular}{l}$118\si{\nano\barn}\text{, Yukawa}$ \\ $122\si{\nano\barn}\text{, Fermi 2P}$%
\end{tabular}
\right. ,
\end{split}
\end{align}
with coefficients given in Table \ref{Tab2} for each distribution. Gaussian and homogeneously charged sphere distributions give results similar to the Fermi 2P distribution. When $\delta=0$, the inequation \eqref{IneqComp} describes a region delimited by an ellipse of which the major axis is parallel to the line $\beta =\alpha $. The effect of first-degree monomials on the ellipse equation is to shift its center and modify the length of the axes. It can be shown that for \eqref{IneqComp}, the translation of the center from the origin due to the interference term is less than $0.2\%$ of the major axis length and the corresponding axis correction is of order $0.001\%$. Therefore, any contribution coming from the interference terms is completely clouded by the theoretical uncertainty and may be neglected.

\begin{table}[]
	\setlength{\tabcolsep}{10pt}
	\renewcommand {\arraystretch}{1.5}
	\centering
	\begin{tabular}{*3c}
		\toprule
		Model                      & With abs.          & Without abs.        \\ 
		\midrule
		Yukawa                     & $42\pm 8\si{\nano\barn}$ & $52\pm 10\si{\nano\barn}$ \\
		Fermi, Gaussian, sphere & $38\pm 8\si{\nano\barn}$ & $45\pm 9\si{\nano\barn}$  \\ 
		\bottomrule
	\end{tabular}
	\caption[SM cross section]{\footnotesize cross section results for the Standard Model calculations of the ATLAS measurement [see Eq. \eqref{Convo}]. The second row shows the results using a Yukawa distribution of charge, while the third row shows the results using Fermi 2P, Gaussian, and homogeneously charged sphere distributions. The second and third column show the results obtained with and without including the absorption (abs.) factor \eqref{FatAbs}. Uncertainties due to the lack of knowledge of the ion's charge distribution are propagated and estimated to be of order $20\%$ of the total cross section \cite{dEnterria}.}
	\label{Tab}
\end{table}

\begin{figure}[h!]
	%\vspace{-5mm}
	\centering
	\includegraphics[scale=0.95]{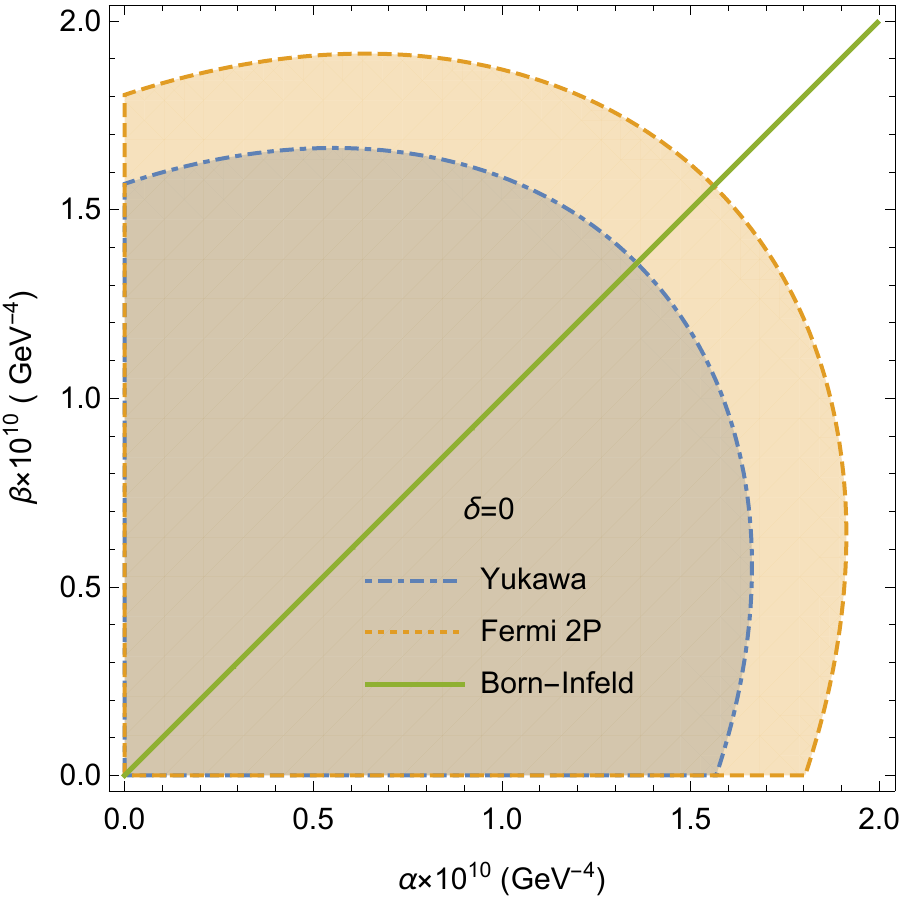}
	%\vspace{-0.6 cm}
	\caption[NL Feynman diagram]{\footnotesize Phase space accessible to the parameters $\alpha $ and $\beta $ derived from \eqref{IneqComp} when $\delta=0$. The more restrictive blue region is obtained with a Yukawa distribution. The broader yellow region is obtained with all three distributions: Fermi with two parameters, Gaussian and homogeneously charged sphere. The green line $\beta =\alpha $ are the values accessible to Born-Infeld-like theories.}
	\label{Vinc1}
\end{figure}

\begin{figure}[h!]
	%\vspace{-5mm}
	\centering
	\includegraphics[scale=0.95]{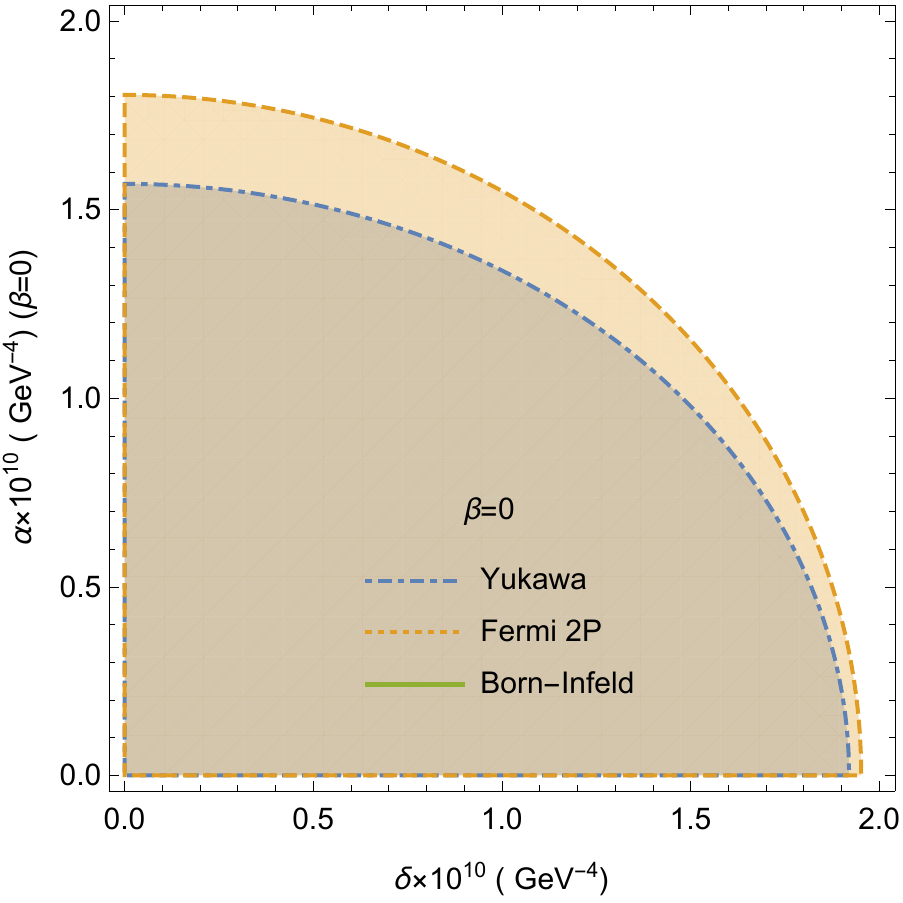}
	%\vspace{-0.6 cm}
	\caption[NL Feynman diagram]{\footnotesize Phase space accessible to the parameters $\alpha $ and $\delta $ derived from \eqref{IneqComp} for the special case when $\beta=0$. Due to symmetry of the inequation, this phase space also corresponds to $\alpha=0$.}
	\label{Vinc2}
\end{figure}

\begin{figure}[h!]
	%\vspace{-5mm}
	\centering
	\includegraphics[scale=0.95]{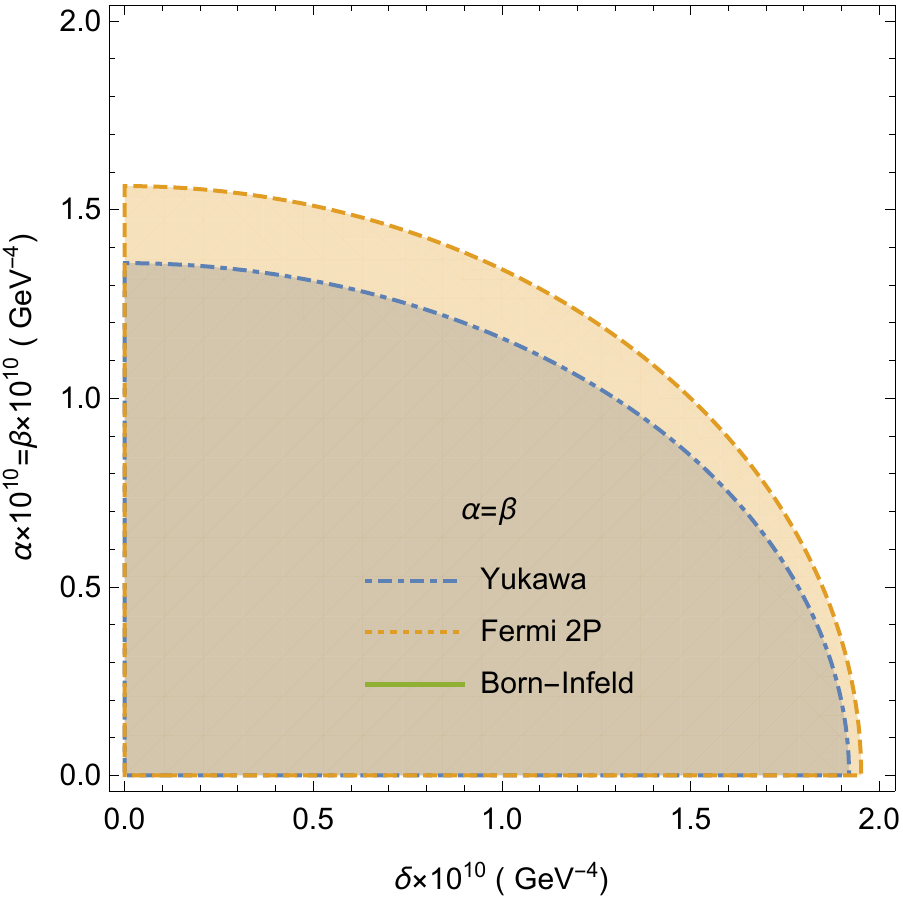}
	%\vspace{-0.6 cm}
	\caption[NL Feynman diagram]{\footnotesize Phase space accessible to the parameters $\alpha $ and $\delta $ derived from \eqref{IneqComp} when $\alpha=\beta$.}
	\label{Vinc3}
\end{figure}

The phase space volume accessible to the parameters $\alpha$ and $\beta$ when $\delta=0$ is presented in Fig. \ref{Vinc1}. We show the outer bounds for Yukawa and Fermi 2P distributions (the Gaussian and homogeneously charged sphere are similar to the latter) as well as the line $\beta =\alpha $ corresponding to Born-Infeld-like theories. As a consequence of the quadratic dependence on the parameters of the cross section due to nonlinear corrections, we are able to completely constrain a finite region of the phase space with one experimental datum. This is not always possible, as is the case of experiments that measure the magnetic birefringence or Lamb shift effect \cite{Fouche}. Additionally, due to causality and unitarity principles, the parameters must be positive \cite{Shabad}. We note that Yukawa distribution is more restrictive than the others. This is due to the fact that it leads to an overestimation of the Standard Model cross section (Table \ref{Tab}), therefore leaving a smaller contribution to nonlinear corrections. As a result, values accessible with the Fermi 2P, Gaussian, and sphere distributions can be up to $15\%$ larger. In Figs. \ref{Vinc2} and \ref{Vinc3} we show the accessible volumes for $\beta=0$ and $\alpha=\beta$ against $\delta$, respectively.

%\newpage

%\newpage

%\newpage

\begin{table}[]
	\setlength{\tabcolsep}{4pt}
	\renewcommand {\arraystretch}{1.5}
	\centering
		\begin{tabular}{*4c}
			\toprule
			{} 								& \multicolumn{2}{c}{$\delta=0$}				& $\beta=0$				\\
			Model 							&  $\alpha =\beta $ 	 & $\beta =0$			& $\delta$				\\
			\midrule
			Yukawa    						& $1.4\times 10^{-10}$   & $1.6\times 10^{-10}$ & $1.9\times 10^{-10}$	\\
			Fermi, Gaussian, sphere		& $1.6\times 10^{-10}$   & $1.8\times 10^{-10}$ & $2.0\times 10^{-10}$		\\
			\bottomrule
		\end{tabular}
	\label{ParSup}
	\caption[Parameter upper values]{\footnotesize Upper values of the parameters for each distribution. The second and third column show the upper values of $\alpha$ and $\beta$ shown in Fig. \ref{Vinc1} for $\delta=0$ when $\alpha =\beta $ and $\beta=0$, respectively. Due to the $\alpha \leftrightarrow \beta $ symmetry of \eqref{IneqComp}, the $\beta=0$ case also corresponds to the upper value of $\beta $ when $\alpha =0$. The third column shows the upper value of $\delta$ shown in Figs. \ref{Vinc2} and \ref{Vinc3}.}
\end{table}

For example, we use the upper limits in order to constrain the parameter of Born-Infeld-like theories defined by $\alpha =\beta =\frac{1}{8b^{2}}$ and $\delta=0$ (see Table \ref{ParSup}). The lower bound obtained using the Yukawa distribution is $b_{Y}\gtrsim 3.0\times 10^{4}\si{\giga\electronvolt}^{2}\simeq 1.3\times 10^{28}\si{\volt\per\meter}$, while for Fermi 2P and the others, it is $b_{F}\gtrsim 2.8\times 10^{4}\si{\giga\electronvolt}^{2}\simeq 1.2\times 10^{28}\si{\volt\per\meter}$. Similarly, we may define a mass $M\equiv \sqrt{b}$, for which we obtain $M\gtrsim 170\si{\giga\electronvolt}$ for all four distributions, in accordance with Ref. \cite{Ellis}.

An upper bound for the $CP$-odd term parameter $\delta$ has been obtained for the first time in the energies accessible to the LHC. From within the Standard Model, contributions to $FG$ are predicted from the weak and strong sectors \cite{Millo2009}.

Finally, with the future project of the ATLAS Collaboration to measure the $\gamma \gamma \rightarrow \gamma \gamma $ scattering with extended tracking acceptance $\left\vert \eta \right\vert <4$ in mind, we calculate the Standard Model prediction using the same remaining cuts to be $\sigma_{Y}=52\pm 10$ nb, with the Yukawa distribution, and $\sigma _{F,G,S}=45\pm 9$ nb with the Fermi 2P, Gaussian and homogeneously charged sphere distributions.

\section{\label{Conclusion} Conclusion}
The recent measurement of the $\gamma \gamma $ scattering by the ATLAS Collaboration has opened a new possibility to test QED and constraining with great precision the phase space where nonlinear corrections live. In this work, using the equivalent photon approximation, we calculate the Standard Model prediction for this phenomenon measured at the ATLAS detector using four nuclear charge distributions (see Table \ref{Tab}). These results are in accordance with the literature \cite{Klusek2016,dEnterria,ATLAS}. We investigated leading-order nonlinear corrections to Maxwell electrodynamics parameterizing the square of the invariants $F^{2}$, $G^{2}$ and $FG$, and obtained an analytic expression for the nonpolarized squared amplitude for the $\gamma\gamma$ scattering \eqref{AmplQuad} as well as for the differential \eqref{SecChoqDif} and total \eqref{SecChoqTot} cross sections.

To constrain the parameters, we deducted the Standard Model cross section prediction from the measured value by the ATLAS Collaboration and interpreted the remaining cross section as coming from the nonlinear corrections. As a consequence of the functional dependence on $\alpha $, $\beta $ and $\delta$, a finite region from the parameters phase space could be completely constrained [Eq. \eqref{IneqComp}]. The interference term between the Standard Model and nonlinear correction amplitudes was analyzed and found to be negligible. Lastly, we have shown the upper bound and its dependence on the nuclear charge distribution (see Figs. \ref{Vinc1}, \ref{Vinc2} and \ref{Vinc3}). 

The constraints obtained in this paper for the nonlinear corrections derived using light-by-light scattering cross section measurement are much more precise than those obtained with any other experiment. When confronted with those obtained in low-energy experiments, as in \cite{Fouche}, our constraints are up to $20$ orders of magnitude lower for $\alpha =\beta $. In \cite{AtomHydr} in which the effects of Born-Infeld-like theories were analyzed using the hydrogen's ionization energy, the lower bound $b\geq 1.07\times 10^{21}\si{\volt\per\meter}$, corresponding to $\alpha=\beta \le8.1\times 10^{4}\si{\giga\electronvolt}^{-4}$, is $14$ orders of magnitude larger. Lastly, $12$ orders of magnitude of precision were obtained when comparing the upper bound for the Born-Infeld parameter in Ref. \cite{Soff}. Also, defining the energy regime in Born-Infeld theory in terms of its parameter as $M\equiv \sqrt{b}$, we obtain the lower bound $M\gtrsim 170\si{\giga\electronvolt}$, which is compatible with Ref. \cite{Ellis}.

A first constraint for the $\delta$ parameter of the $CP$-odd term $FG$ of the order of $\delta\sim10^{-10}\si{\giga\electronvolt}^{-4}$ was obtained in the LHC energy scale. Although from a different energy regime, an estimation of the contribution from the strong sector in the optical energies has been calculated using the chiral perturbation theory with a $\theta$-parameter of the order of $\theta\sim10^{-10}$ to be $\delta\sim10^{-15}\si{\giga\electronvolt}^{-4}$ in Ref. \cite{Millo2009}.

The ATLAS Collaboration is aiming to improve the tracking acceptance from $\left\vert \eta \right\vert <2.5$ to $\left\vert \eta \right\vert <4$. With this in mind, we calculated the cross section to be measured as $45\pm 9$ nb using the realistic Fermi with two parameters nuclear charge distribution.

The first direct observation of the $\gamma\gamma $ scattering made by the ATLAS Collaboration is, without any doubt, a great achievement. This mechanism proves to be an elegant and efficient way to probe the quantum vacuum which allows constraining a great variety of beyond Standard Model theories. As a matter of fact, LHC p-p and Pb-Pb UPC measurements have been used to bound the axion-like particles-photon coupling constant for axion-like particles masses above $1\si{\giga\electronvolt}$ \cite{Knapen,KnapenConf,Baldenegro}. While this first measurement is compatible with QED predictions, its $40\%$ absolute uncertainty is still an obstacle to overcome. Future measurements, with greater precision, wider phase space, and higher-energy regimes, will allow us to analyze with greater detail several contributions that compose the mechanism. As commented in Ref. \cite{Klusek2011}, forthcoming light-by-light scattering measurements could be used to constrain nuclear charge distributions.

Finally, the increasing energy scales and experiment precision will impose more sophisticated theoretical analysis. In the scope of nonlinear corrections to Maxwell electrodynamics, in order to obtain more precise constraints in these scenarios, a future investigation would be to include higher-order terms from the Lagrangian expansion.

\section*{Acknowledgments}
The authors are thankful to Thomas Hahn for his valuable help with \texttt{FeynArts}, \texttt{FormCalc}, and \texttt{LoopTools} packages. They are also grateful for CNPq-Brazil's financial support.

\bibliography{Bibliography}
\end{document}